# A new paradox in random-effects meta-analysis

Jiandong Shi, MSc[1]; Aimin Wu, MD[2]; Tiejun Tong, PhD[1]

[1]Department of Mathematics, Hong Kong Baptist University, Hong Kong

[2]Department of Orthopaedics, The Second Affiliated Hospital and Yuying Children's Hospital of Wenzhou Medical University, Zhejiang Provincial Key Laboratory of Orthopaedics, Wenzhou, China

Corresponding Author: Tiejun Tong, Department of Mathematics, Hong Kong Baptist University, Hong Kong (tongt@hkbu.edu.hk)

**Abstract**

Meta-analysis is an important tool for combining results from multiple studies and has been widely used in evidence-based medicine for several decades. This paper reports, for the first time, an interesting and valuable paradox in random-effects meta-analysis that is likely to occur when the number of studies is small and/or the heterogeneity is large. With the incredible paradox, we hence advocate meta-analysts to be extremely cautious when interpreting the final results from the random-effects meta-analysis. And more importantly, with the unexpected dilemma in making decisions, the new paradox has raised an open question whether the current random-effects model is reasonable and tenable for meta-analysis, or it needs to be abandoned or further improved to some extent.

Meta-analysis is an important tool for combining results from multiple studies and has been widely used in evidence-based medicine for several decades. The evidence from meta-analysis is nowadays regarded as the strongest evidence in medical practice. The main purpose of meta-analysis is to collect summary data from similar studies in an effort to increase the statistical power and, consequently, obtain more reliable results than those from each individual study. In other words, we are often dealing with scenarios as in Figure 1, where the effect sizes are statistically significant only in some of the studies. By performing a meta-analysis, we are able to aggregate information from all included studies and, with the increased power, the overall effect can be detected as significant.

Following the spirit of meta-analysis, it would then be hard to imagine the opposite scenario may also occur. For instance, if a new drug is significantly better than a placebo in each study, by meta-analysis we conclude, however, that there is no significant difference between the two treatments. To explain why such a scenario is incredible, one may refer to, for example, the claim by Jacob Stegenga who is a philosopher of science at the University of Cambridge. In a recent *Science* article[1] entitled "The Metawars", the author quoted his words as follows: *"When the evidence points clearly in one direction, there is little need for a meta-analysis."*

**A new paradox in random-effects meta-analysis**

When reading the above *Science* paper, we suddenly realized that Stegenga's claim is indeed not correct, but rather yields an interesting and valuable paradox in meta-analysis. To verify the existence of the new paradox, we revisited our recent paper[2] on meta-analysis of fusion surgery for lumbar spinal stenosis. As displayed in Figure 2, with the 95% confidence intervals (CIs) of the difference of two hospital stays as [0.19, 0.71] and [1.17, 2.34], both studies[3,4] showed that the hospital stay of the experimental group was significantly longer than that of the control group. Hence, if Stegenga's claim is followed, one would naturally conclude that the overall comparison will remain the same as in each study. While on the other side, with the heterogeneity statistic ($I^2$) at 94%, by the Cochrane Handbook[5] the random-effects model was chosen to synthesize the results from the two studies. Surprisingly, the overall effect from the random-effects model had a 95% CI of [-0.20, 2.35]; that is, the meta-analytic results did not show any significant difference between the two lengths of hospital stay. This example, therefore, confirms our conjecture that the new paradox is real.

In accordance with the above findings, we state our newly discovered paradox as follows:

*"For meta-analysis of continuous outcomes, assume that the individual effect sizes are all significantly larger (or smaller) than zero. A paradox occurs if the overall effect from the random-effects model is, however, not significantly larger (or smaller) than zero."*

Accordingly, for meta-analysis of binary outcomes, a similar statement for the paradox can be made that compares the effect sizes to one rather than zero. All other arguments remain the same as those for continuous outcomes. To the best of our knowledge, this incredible paradox has never been reported in the literature on meta-analysis.

**Other paradox examples in the literature**

With the newly discovered paradox, we examined many meta-analysis papers in the literature and found that the paradox occurs more frequently than we expected. In particular, when the number of studies is small, the occurrence of the paradox tends to be considerably large. It is noteworthy that the new paradox also appears in leading medical journals, and for illustration, we present 3 typical examples in what follows, with one[6] in *BMJ*, one[7] in *Lancet*, and the other[8] in *JAMA*.

The paradox example[6] in *BMJ* is on binary outcomes measured with odds ratios, where the authors were to explore the association of the dipeptidyl peptidase-4 inhibitors and risk of heart failure in type 2 diabetes. It can be seen in Figure 3 that, with the 95% CIs of two individual odds ratios as [1.04, 1.41] and [1.16, 2.92], both studies[9,10] showed a significantly increased risk of admission for heart failure in patients treated with DPP-4 inhibitors versus no use. According to Stegenga's claim, the use of DPP-4 inhibitors should be avoided in the clinical treatments of type 2 diabetes. On the other side, with the heterogeneity statistic ($I^2$) at 65%, the authors applied the random-effects meta-analysis and so made a different conclusion as the 95% CI of the overall effect is [0.95, 2.09]. This example shows that the paradox may also occur in meta-analysis of binary outcomes.

The article[7] in *Lancet* serves as another good example for illustrating the paradox with binary outcomes, in which the odds ratio was used as the effect size to investigate the prevalence and risk of violence against adults with mental illness. As shown in Figure 4, with the 95% CIs of the individual odds ratios as [1.44, 4.04], [1.46, 2.69] and [10.01, 13.92], all three studies[11,12,13] suggested that the adults with mental illness had a significantly higher risk of suffering from the violence than the normal ones. Then if we follow Stegenga's claim, the overall result of the higher prevalence and risk of violence against adults with mental illness can be naturally conducted. However, with the heterogeneity statistic ($I^2$) at 99%, the random-effects model was applied and the overall effect had a 95% CI of [0.91, 16.43], which disagreed with our intuition as Stegenga claimed.

In addition, we also found one more paradox example[8] with continuous outcomes in *JAMA*. The study was to explore the association of the use of plant-based therapies and menopausal symptoms. As displayed in Figure 5, with the 95% CIs of the difference of frequency of night sweat as [-0.79, -0.01] and [-4.5, -3.3], both studies[14,15] showed that the experimental group had a significantly decreased frequency of night sweat in 24 hours compared with the control group. And hence if we follow Stegenga's claim, the use of plant-based therapies shall be recommended automatically. Nevertheless, with the heterogeneity statistic ($I^2$) at 99%, the random-effects model was applied and it provided a different conclusion with the 95% CI of the overall effect being [-5.57, 1.29].

**Further discussion**

With the above paradox examples, we are now aware that the new paradox may cause great confusion in medical practice. Given that the meta-analytic evidence differs from the unanimous evidence from all individual studies, how should we interpret the final results and be certain of the final claims? Taking the fusion surgery study[2] as an example, if we follow the common practice that the meta-analytic evidence is at the peak of the evidence pyramid, then we would draw the conclusion of no significant difference between the two lengths of hospital stay. Yet on the other side, both studies[3,4] were published in the prestigious journal *NEJM*, and they were of very high quality and reliable themselves. Hence, certain researchers would be more willing to accept the viewpoint delivered in each paper that the lengths of hospital stay are in fact different. With the above contradictory viewpoints, the new paradox has been putting us in a dilemma on what final conclusion can be made, and/or whether meta-analysis should be routinely performed.

To explain why the new paradox occurs, we have also been working hard on the statistical aspects to deeply expose the underlying reasons behind the paradox. Due to space limit, we will not be able to unfold all the theoretical results in this short report but rather present three interesting remarks. First, our new paradox is different from the well-known paradoxes including Simpson's paradox, and so is discovered for the first time in the literature. Second, the new paradox is likely to occur when the number of studies is relatively small, say 2 or 3 studies, as demonstrated in the above paradox examples. Third, the new paradox is likely to occur when the heterogeneity is large; while for the fixed-effect model with no heterogeneity, the new paradox will never occur. Specifically for the fixed-effect model, it can be shown that the variance of the overall effect is smaller than every individual variance. Thus by noting that the overall effect is a weighted mean of all individual effect sizes, if the individual effect sizes are all positively (or negatively) significant, the confidence interval of the overall effect will not cross the zero vertical line. Consequently, there is no paradox in the fixed-effect model.

**Concluding remarks**

With the incredible paradox, we hence advocate meta-analysts to be extremely cautious when interpreting the final results from the random-effects meta-analysis, especially when the number of studies is small and/or the heterogeneity is large. It is also noteworthy that meta-analyses with few studies are very common in the literature. According to a recent study on the Cochrane Database of Systematic Reviews[16], the median number of studies per meta-analysis is only 3 among a total of 22,453 health-related meta-analyses. In other words, at least half of meta-analyses conducted in the literature are with only 2 or 3 studies so that the paradox may have a considerably large chance to occur.

To conclude, we reiterate that the new paradox has been putting us in a dilemma on what final medical conclusion can be made. And meanwhile, the new paradox has raised an open question whether the current random-effects model is reasonable and tenable for meta-analysis, or it needs to be abandoned or further improved to some extent.

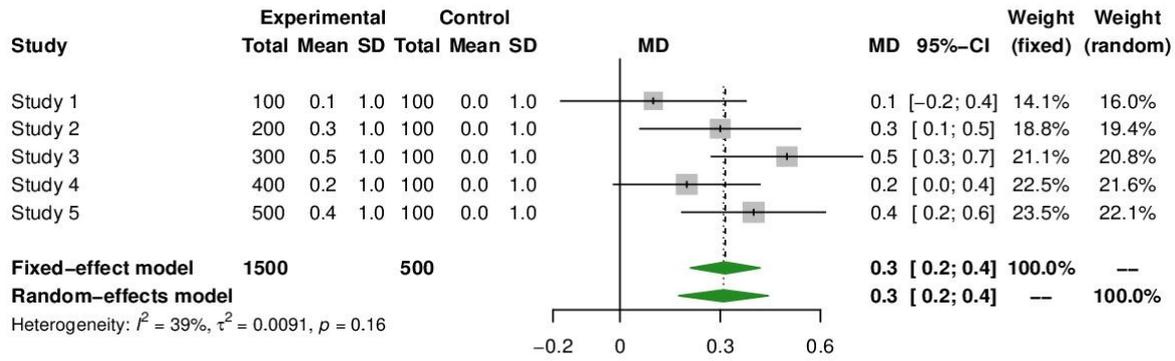

figure 1: A typical example of meta-analysis with five hypothetical studies

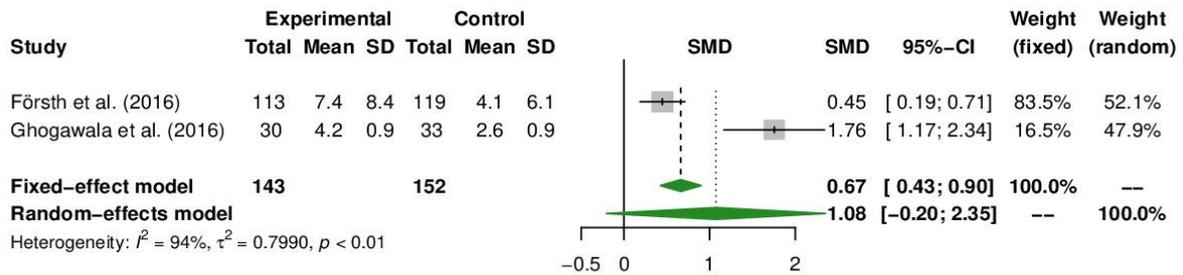

Figure 2: The paradox in Wu et al. (2016)

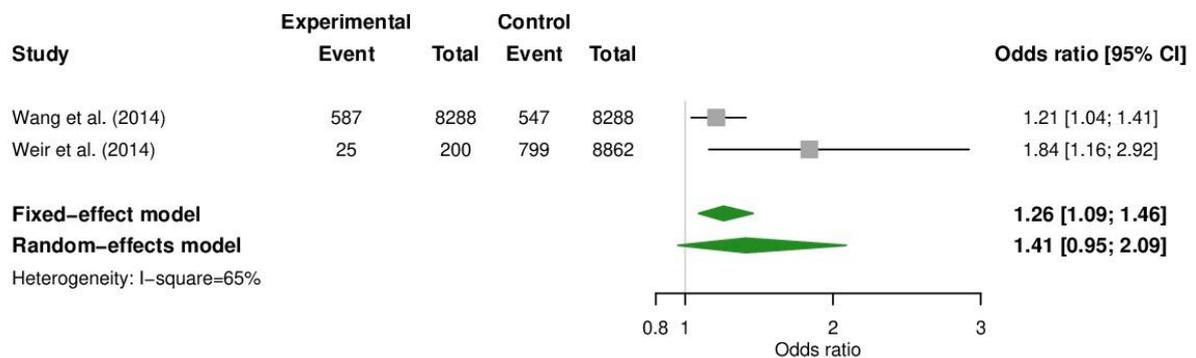

Figure 3: The paradox in Li et al. (2016)

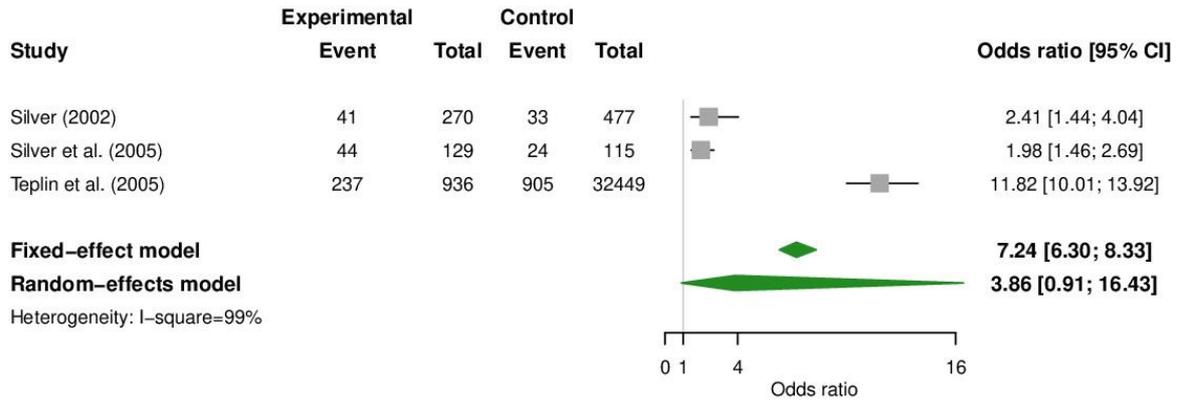

Figure 4: The paradox in Jones et al. (2012)

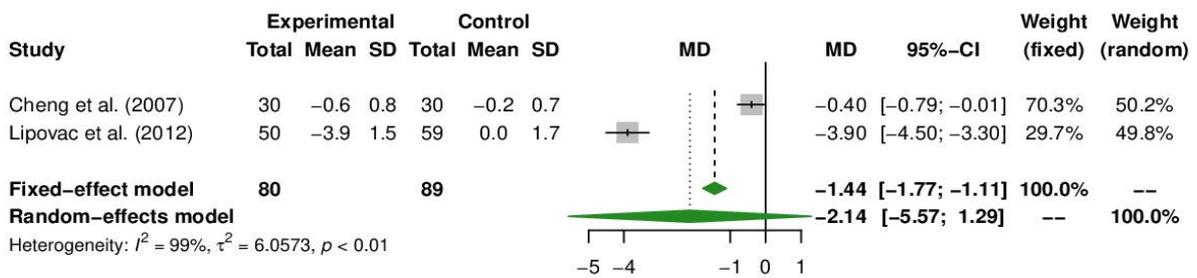

Figure 5: The paradox in Franco et al. (2016)